\documentclass[traditabstract]{aa} 
\usepackage{amsmath}
\usepackage{graphicx}
\usepackage{txfonts}
\usepackage{comment}
\usepackage{natbib}
\bibpunct{(}{)}{;}{a}{}{,}
\usepackage{float}
\usepackage{calc}
\usepackage{textcomp}
\usepackage{placeins}
\usepackage{color}
\usepackage{rotating} 
\usepackage{lscape}
\usepackage{url}
\usepackage[colorlinks=true,linkcolor=blue,citecolor=blue]{hyperref}
\usepackage{subfig}
\usepackage[all]{draftcopy}
\usepackage{longtable}
\usepackage{array}
\usepackage{threeparttable}
\usepackage{lscape}
\usepackage{enumerate}
\DeclareTextSymbol{\degre}{OT1}{23}

\newcounter{savedfootnote}

\def \cigale{{{\sc cigale}}}
\renewcommand{\epsilon}{\varepsilon} 

\begin{document}
\title{Dust emission from the bulk of galaxies at the Epoch of Reionization}

\author{
L.~Ciesla\inst{1}\fnmsep\thanks{\email{laure.ciesla@lam.fr}},
S.~Adscheid\inst{2},
B.~Magnelli\inst{3},
M.~Boquien\inst{4},
N.~Laporte\inst{1},
M.~B\'ethermin\inst{5},
C.~Carvajal\inst{1},
E.~Schinnerer\inst{6},
and D.~Liu\inst{7,8}.
}

\institute{	
Aix Marseille Univ, CNRS, CNES, LAM, Marseille, France
\and
Argelander-Institut f\"ur Astronomie, Universit\"at Bonn, Auf dem H\"ugel 71, 53121 Bonn, Germany
\and
Université Paris-Saclay, Université Paris Cité, CEA, CNRS, AIM, 91191, Gif-sur-Yvette, France
\and
Université Côte d'Azur, Observatoire de la Côte d'Azur, CNRS, Laboratoire Lagrange, 06000, Nice, France
\and 
Universit\'e de Strasbourg, CNRS, Observatoire astronomique de Strasbourg, UMR 7550, 67000 Strasbourg, France
\and
Max Planck Institute for Astronomy, Konigstuhl 17, 69117 Heidelberg, Germany
\and
Max-Planck-Institut fur Extraterrestische Physik (MPE), Giessenbachstr., 85748, Garching, Germany
\and
Purple Mountain Observatory, Chinese Academy of Sciences, 10 Yuanhua Road, Nanjing 210023, People's Republic of China
}		

   \date{Received; accepted}

  \abstract
{
The excess of UV bright galaxies observed at $z>10$ has been one of the major surprises from the JWST early observations. 
Several explanations have been proposed to understand the mild change in space density of the UV bright galaxies at these high redshifts, among them an evolution of dust attenuation properties in galaxies. 
However, our view of dust in primordial galaxies is limited towards a few tens of $z\sim7$ galaxies, pre-selected from UV-optical observations, and are thus not necessarily representative of the bulk of the sources at these redshifts. 
In this work, we aim at constraining the dust properties of galaxies at $6<z<12$ by making the most of the A$^3$COSMOS database in the JADES/GOODS-South field.
We stacked ALMA band 6 and 7 observations of 4464 JADES galaxies covered by the A$^3$COSMOS database and used the measurements as constraints to perform UV-to-FIR SED modelling.
We obtain tentative signals for the brightest UV galaxies ($M_{\mathrm{UV}}<-19$~mag) as well as for the most massive ones ($\log M_\star/M_\odot>9$) at $6<z<7$, and upper limits for fainter ($M_{\mathrm{UV}}>-19$~mag), lower mass sources ($\log M_\star/M_\odot<9$), and at higher redshift ($z>7$).
Fitting these $6<z<7$ galaxies with ALMA constraints results in lower star formation rates ($-0.4$\,dex) and FUV attenuation ($-0.5$\,mag) for galaxies with $\log M_\star/M_\odot>8$, compared to the fit without FIR.
We extend the $L_{\mathrm{IR}}$ vs $M_{\mathrm{UV}}$ relation down to $M_{\mathrm{UV}}=-19$~mag and show a tentative breakdown of the relation at fainter UV magnitudes.
The positions of the JADES $z\sim6.5$ sample on the infrared excess (IRX) versus $\beta$ and IRX versus $M_\star$ diagrams are consistent with those of the ALPINE ($z\sim5.5$) and REBELS ($z\sim6.5$) samples, suggesting that the dust composition and content of our mass-selected sample are similar to these UV-selected galaxies.
Extending our analysis of the infrared properties to $z>7$ galaxies, we find a non-evolution of $\beta$ in the $M_{\mathrm{UV}}$ range probed by our sample (-17.24$^{+0.54}_{-0.62}$) and highlight the fact that samples from the literature are not representative of the bulk of galaxy populations at $z>6$.
We confirm a linear relation between A$_{\rm V}$ and sSFR$^{-1}$ with a flatter slope than previously reported due to the use of ALMA constraints.
Our results suggest that rapid and significant dust production has already happened by $z\sim7$.
}

   \authorrunning{Ciesla et al.}
   \titlerunning{Dust emission at the Epoch of Reionisation}

   \maketitle

\section{\label{intro}Introduction}

In the interstellar medium (ISM) of galaxies, dust plays a crucial role, catalysing the transformation of atomic hydrogen into molecular hydrogen, the raw material for star formation \citep{Wolfire95}. 
Dust also enables gas to cool and condense into new stars by absorbing ultraviolet (UV) radiation from nearby young stars \citep{Draine78, Dwek86, Hollenbach&Tielens97}. 
This absorbed energy is then re-emitted in the infrared (IR) range, where dust grain thermal emission dominates the spectral energy distribution (SED) of galaxies, covering wavelengths from approximately 8 to 1000\,$\mu$m. 
Because of its important role in the ISM and its tight link with the other components of galaxies, the study of dust emission is paramount for a better understanding of all of the processes at play, especially galaxies' star formation activity.

At $z>4$, most of the constraints on star formation in galaxies are obtained via UV emission which is highly sensitive to dust attenuation and needs to be corrected to obtain a complete view \citep[e.g.,][]{Calzetti00,SalimNarayanan20}.
Understanding how dust attenuation varies with redshift and other galaxy properties is thus essential for accurately determining star formation rates (SFRs), and thereby constructing a precise account of galaxy evolution throughout cosmic history.
At $z>4$, two programs targeted UV-selected galaxies to be observed with the Atacama Large Millimeter/Submillimeter Array (ALMA). 
ALPINE \citep[ALMA Large Program to INvestigate CII at Early times][]{LeFevre20,Bethermin20,Faisst20} provided 23 far-infrared continuum-detected galaxies at $z\sim5$. 
At slightly higher redshift, the REBELS survey \citep[Reionization Era Bright Emission Line Survey,][]{Bouwens22} provided dust continuum ALMA observations of 40 UV-selected bright galaxies at $6.5<z<7.7$.
At $z>7$, a few galaxies have been detected in dust emission \citep[e.g.,][]{Watson15,Laporte17,Laporte19,Bowler18,Tamura19,Bakx20,Sugahara21,Endsley22,Witstok22,Schouws22,Algera23,Schouws24} providing hints that significant dust reservoirs are already in place at early times.
Although they provide an indispensable and detailed view on dust emission at $z\gtrsim4$, these UV-selected samples are limited in terms of statistics and might only represent the tip of the infrared luminosity distribution at these redshifts.
To overcome this, with a different approach, \cite{Magnelli24} used the A$^3$COSMOS ALMA database \citep{Liu19,Adscheid24} gathering the wealth of ALMA archival observations to statistically probe dust properties of a mass selected sample at $4<z<5$.

The infrared excess (IRX$\equiv L_{\mathrm{IR}}/L_{\mathrm{UV}}$) vs $\beta$ relation is widely used as a dust attenuation diagnostic \citep{Calzetti94, Meurer99,Calzetti00}.
While this relation provides accurate estimates of the attenuation for local starburst galaxies \citep{Meurer99}, the position of normal star-forming galaxies relative to it is still debated \citep[e.g.,][]{Capak15, Barisic17, Hashimoto19, Bakx20, Bowler22, Schouws22, Boquien22, Magnelli24, Bowler24}.
The position of galaxies with respect to the IRX-$\beta$ relation depends on the geometry of dust and stars in the galaxy \citep{Popping17,Narayanan18,Ferrara22, Pallottini22,Vijayan22} and the effect of the star formation history (SFH) is also expected since the UV $\beta$ slope is a combination of dust attenuation and the intrinsic UV slope, $\beta_0$ \citep{Boquien12}.
\cite{Magnelli24} used averaged measurements of ALMA galaxies using A$^3$COSMOS to investigate the dust attenuation properties of star-forming galaxies at $4<z<5$.
They find that the average properties of their sample are compatible with the relations of \cite{Meurer99} and \cite{Calzetti00} for local starbursts. 
Similar conclusions were reached by \cite{Bowler24} on stacked properties of the REBELS galaxies at z$\sim$7.
Both results point towards non-evolving grain properties (e.g., size distribution, composition).
However, using the ALPINE sample of galaxies at $4.4<z<5.5$, \cite{Boquien22} found a wider variety of attenuation curves among these sources from relations steeper than the Small Magellanic Cloud (SMC) extinction curve to relations shallower than for local starbursts.
In their fitting process, the slope of the attenuation curve, $\delta$, was let free which could explain the larger variety of curves that they find.

The relation between IRX and the stellar mass is another diagnostic of the dust attenuation properties, with the IRX quantifying the dust obscured fraction while the stellar mass is a proxy of the past star formation activity, hence of dust production. 
Recently, there have been claims of a steepening of the relation between IRX and stellar mass with increasing redshifts \citep{Fudamoto20,Magnelli24}, that is a decreasing obscured fraction with redshifts.
This would imply an evolution of dust geometry relative to new-formed stars.
Using the REBELS sample of galaxies at $z\sim 7$, \cite{Bowler24} also found a decrease in the obscured fraction with redshifts but associated with a IRX-stellar mass relation with a relatively flat slope, in contradiction with the results of \cite{Fudamoto20} and \cite{Magnelli24}, thus adding to the debate.

The goal of this work is to extend these studies in terms of statistics but also redshift with a statistical analysis using the A$^3$COSMOS (A$^3$GOODSS) archival data, pushing it to its limits by stacking galaxies at $z=6$ and beyond, observed as part of the JADES survey \citep{Rieke23}.
Indeed, dust at higher redshift ($z\gtrsim9-10$) is also central in the discussion to understand the origin of the excess of UV bright galaxies observed at $z>10$ \citep[e.g.,][]{Castellano22,Naidu22,Donnan23,Harikane23,Arrabal23,CurtisLake23,Robertson23,Harikane23b,RobertsBorsani24}. The evolution of attenuation properties is mentioned as a possible explanation to the abundance of such ``blue monsters'' \citep[e.g.,][]{Ferrara23}.
Constraints on the infrared emission of early galaxies could provide benchmark properties for these studies.

The paper is organised as follows: in Sect.~\ref{sample} we describe the JADES sample used as prior for the ALMA stacking and in Sect.~\ref{sec:sedfit} the SED fitting procedure to obtain the UV magnitudes and stellar mass from which we bin them. The ALMA stacking of the sources using A$^3$COSMOS is detailed in Sect.~\ref{stack}. We compare the properties of the galaxies with and without using the ALMA constraint in the SED modelling in Sect.~\ref{impact} and study the dust properties of the sample in Sect.~\ref{dust_results}. Constraints on galaxies at $z>7$ are provided in Sect.~\ref{highz}.
Throughout the paper, we use a \cite{Salpeter55} initial mass function and WMAP7 cosmology \citep{Komatsu11}.
\section{\label{sample}The sample}

\subsection{Sample selection}
We use the second release of the JWST Advanced Deep Extragalactic Survey \citep[JADES,][]{Rieke23,Bunker23,Eisenstein23,Hainline23} of the Hubble Ultra Deep Field covering 68\,sq.arcmin presented in \cite{Eisenstein23b}.
Although the main part of the study focuses on galaxies with $6<z<7$, we select galaxies with $6<z<12$ in the JADES catalogue.
We refer to \cite{Rieke23}, \cite{Eisenstein23b}, and \cite{Robertson24} for more details on the data reduction process, photometry procedures, and redshift estimates.
This catalogue combines nine broad and five medium JWST/NIRCam bands from the JADES and JEMS \citep{Williams23} surveys, respectively, and existing HST imaging, for a total of 23 photometric bands.
The reddest NIRCam filter, where the galaxies were selected (F444W), allows us to probe the 0.63\,$\mu$m and 0.34\,$\mu$m rest frame at $z=6$ and $z=12$, respectively.
In this work we use the photometry catalogue made from KRON apertures.
In the broad NIRCam bands, the 5$\sigma$ flux depths are between 1.9 and 3.0\,nJy, while they range from 1.6 to 3.3\,nJy in the medium bands.
Photometric redshifts available in the JADES catalogue are computed with the code \texttt{EAZY} \citep{Brammer08}.
The average offset between photometric redshifts and a compilation of spectroscopic redshifts is 0.05 with a $\Delta z/(1+z)$ of 0.024.
We use spectroscopic redshifts when available although it concerns only 0.6$\%$ of the sample at $6<z<7$, and 0\% at $z>7$.
We also checked for spectroscopic data through the Dawn JWST Archive\footnote{\url{https://dawn-cph.github.io/dja/}} and find 37 NIRSpec spectra.
Out of these, 9 of them are galaxies with $z_{spec}<5.5$, we exclude them from our analysis. 

As in \cite{Ciesla24}, we impose each galaxy to be detected with a signal-to-noise (S/N) higher 3 in at least 3 JWST filters. 
This criterion results in a subsample of 8810 galaxies with redshifts between 6 and 12, 6000 of them between 6 and 7.
Given the uncertainty on photometric redshift, several studies impose a criterion on the quality of the photometric redshift, based on its PDF \citep[e.g.,][]{Cole23,Finkelstein23}.
However, in the following, we consider the full galaxy sample.
Indeed after redoing the analysis of this work imposing the same photo-$z$ quality criteria as \cite{Cole23}, we find no strong impact on our results, a conclusion also reached by \cite{Cole23}.

A particular focus on the $6<z<7$ bin is made in the paper, driven by the tentative signal obtained from stacking the ALMA observations that is more constraining at $6<z<7$ than at higher redshifts.
We extend our analysis at higher redshifts in Sect.~\ref{highz}.

\subsection{\label{sec:sedfit}SED modelling with \texttt{CIGALE} and physical properties}

We use the SED modelling code \texttt{CIGALE}\footnote{\url{http://cigale.lam.fr}} \citep[Code Investigating GALaxy Emission,][]{Boquien19} on the HST+JWST set of data, probing the UV to NIR restframe wavelength range.
The fitting set up of \texttt{CIGALE} is the same than the one tested and used in \cite{Ciesla24} which adopts a non-parametric star formation history (SFH) specifically tailored for early galaxies.
The non-parametric approach usually assumes a prior to handle the relation between two consecutive SFR bins of the SFH (continuity, bursty continuity, etc.).
In this work, we use the non-parametric SFH from \cite{Ciesla24} who used no prior and showed that it provides more flexibility to model the bursty SFH expected in early galaxies.
We note that this SFH library results in a wide range of $\beta_0$, the intrinsic UV slope, ranging from -2.78 to 2.65 with a median value of -2.20.
We use the stellar population models of \cite{BruzualCharlot03}, and a \cite{Calzetti00} dust attenuation law which is found to be suitable for galaxies at $z>6$ \citep{Bowler22,Markov24}.

\texttt{CIGALE} assumes energy balance: the energy absorbed in the UV-optical is re-emitted in the IR.
Some studies have shown a spatial offset between the stellar (UV) and dust (IR) emission in some high redshift sub-millimetre galaxies \citep[e.g.,][]{Hodge16,Elbaz18} resulting in discussions on the validity of UV-to-IR SED modelling based on energy balance.
Using the disturbed morphology of M\,82 as a test-bed, \cite{Seille22} showed that the estimates of the star formation rate and stellar mass for the whole galaxy are found to be consistent with the sum of the same parameters obtained for individual regions.
The same conclusion is reached by \cite{Li24} at $4<z<6$ from a resolved analysis of ALMA-CRISTAL galaxies therefore suggesting that there is no bias due to outshining by young stars on the derived global properties of these sources.
Furthermore, using FIRE zoom-in simulations of dusty and high-redshift galaxies, \cite{Haskell23} demonstrated that when the quality of the fit is acceptable, the fidelity of SED modelling estimates using energy balance is independent from the degree of UV/FIR offset, with similar quality than previously reported for local galaxies.
Therefore, we are confident in using the ALMA constraint and perform the SED modelling from UV to FIR using energy balance.

To model the IR emission, we use the dust library of \cite{Schreiber18} that has two free parameters.
One the fraction of dust mass locked in the form of Polycyclic Aromatic Hydrocarbons (PAH), which can vary from 0 to 100\%.
The second is the dust temperature, varying from 15 to 60\,K.
The data used in this work does not allow to constrain the PAH fraction.
Therefore we choose to fix this input parameter to 3\% which corresponds to the estimate of \cite{Schreiber18} at $z\sim3$, the highest redshift constraint provided in their study.
Regarding dust temperature, following the results of \cite{Sommovigo22} from REBELS galaxies, we set it to 47\,K and test in Sect.~\ref{tdust} the impact of this dust temperature assumption on our results.

\begin{table*}[!htbp]
   \centering
   \caption{\cigale\ input parameters used to fit the JADES sample. This set of parameters resulted in 256000 SED models per redshift.}
   \begin{tabular}{l c l}
   \hline\hline
   \textbf{Parameter} & \textbf{Value} & \textbf{Definition}\\
   \hline
   \multicolumn{3}{l}{\textbf{Non-parametric SFH --} \sc{sfhNlevels$\_$flat}}\\[1mm]  
   $age$ (Myr) & 200, 300, 500, 750& maximum possible age\\
   sfr$\_$max & 500 M$_{\odot}$yr$^{-1}$ & max. value of the SFR \\[1mm]
   $N_{\rm bins}$ & 10 & \# of time bins of the SFH \\[1mm]
   $N_{\rm SFH}$ & 500 & \# of SFH per $age$ value \\[1mm]
   Last bin size & 10\,Myr & \\[1mm]
   \multicolumn{3}{l}{\textbf{Stellar population --} \sc{bc03}}\\[1mm]  
   metallicity      &  0.004, 0.02     & \\   
   \multicolumn{3}{l}{\textbf{Emission lines --} \sc{nebular}}\\[1mm]  
   $\log$U      &  -3, -2     & ionisation parameter\\
   $f_{\rm esc}$      &  0, 0.1     & Lyman continuum photons escape fraction\\
   $f_{\rm dust}$      &  0, 0.1     & Lyman continuum photons fraction absorbed by dust\\
   \multicolumn{3}{l}{\textbf{Dust attenuation --} \sc{dustatt\_modified\_starburst}}\\[1mm]  
   E(B-V) lines      &  $\big[$0;\ 0.5$\big]$     &8 values linearly sampled  \\
   E(B-V)s factor      &  0.5     & colour excess ratio between continuum \& nebular  \\
   \multicolumn{3}{l}{\textbf{Dust emission --} \sc{schreiber2016}}\\[1mm] 
   $f_{\rm PAH}$      &  3\%     & fraction of PAH  \\
   $T_{\rm dust}$     &  47\,K    & dust temperature  \\
   \hline
   \label{inputparam}
   \end{tabular}
\end{table*}

To estimate the physical properties, the models are fitted to the observations and the likelihoods are computed from the $\chi^2$. 
The galaxy properties are taken as the likelihood-weighted means and the uncertainties are the likelihood-weighted standard deviations. 
One important aspect to note is that upper limits are taken into account in the fitting procedure \citep[see Sect.~4.3 of][for the presentation of the mathemical implementation]{Boquien19}.
In this work, we take 3$\sigma$ observational uncertainties as the upper limits.

\begin{figure}[!htbp] 
 	\includegraphics[width=\columnwidth]{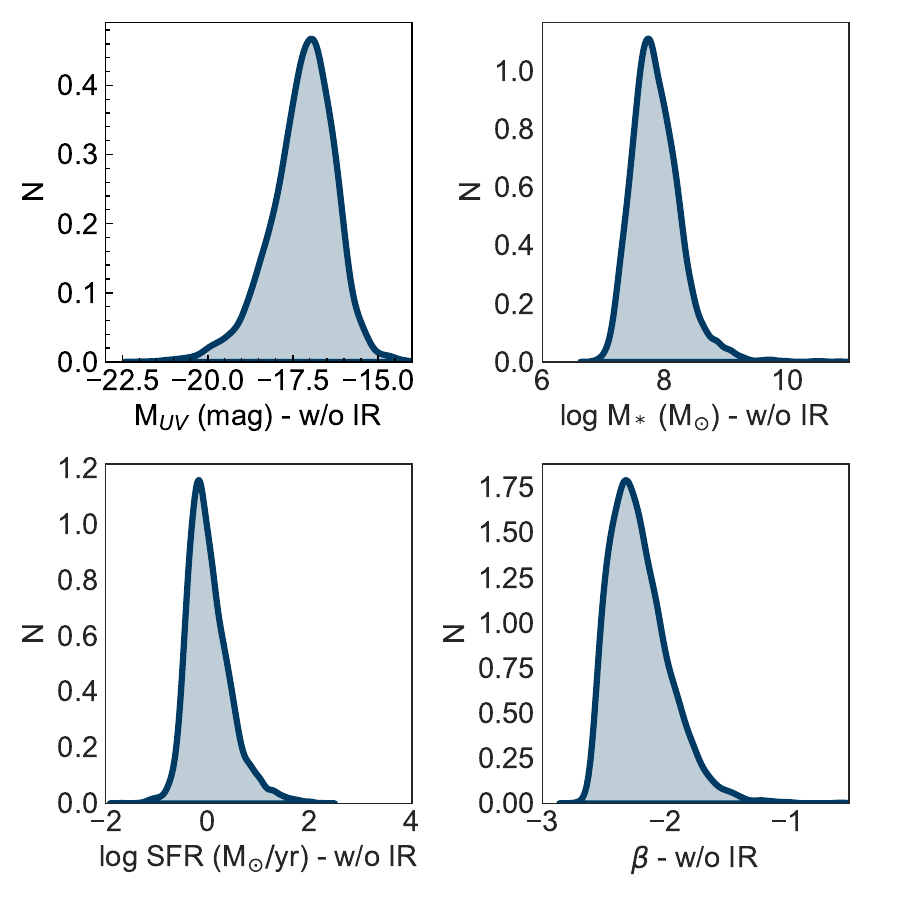}
  	\caption{\label{fig:stats_noir} Normalised distributions of the physical properties of the $6<z<7$ JADES selected sample obtained from UV-to-NIR rest-frame SED modelling (without IR). From left to right, and top to bottom: UV magnitude, stellar mass, star formation rate, and UV slope $\beta$.}
\end{figure}

The parameters used for the fit are provided in Table~\ref{inputparam}.
In their analysis, \cite{Ciesla24} showed that the physical properties, such as stellar mass and SFR, of galaxies with $6<z<12$ are well constrained with the HST+JWST set of filters of the JADES survey.
As we are using the same combination of filters, we rely on their results.
We first apply this \texttt{CIGALE} setup on the JADES sample using only HST and JWST observations.
The resulting normalised distributions of UV magnitude, stellar mass, SFR, and UV slope ($\beta$) are shown in Fig.~\ref{fig:stats_noir} for the $6<z<7$ galaxies.
Our sample allows us to probe relatively faint UV magnitude (-17.11$^{0.55}_{0.63}$\,mag), a median stellar mass of $\log M_\star/M_\odot=7.82^{0.27}_{0.22}$, median SFR of $\log SFR/M_\odot.yr^{-1}=-0.05^{0.30}_{0.21}$, and relatively blue UV $\beta$ slopes (-2.24$^{0.18}_{0.14}$).
The sample that we use for the ALMA stacking is built from this HST+JWST \texttt{CIGALE} run.
We select galaxies with a reduced $\chi^2$ no larger than 4, removing 26 galaxies among which many sources with SEDs that are similar to brown dwarves' SEDs.
The second run of CIGALE using the ALMA constraints is presented in Sect.~\ref{impact}.

\section{\label{stack}ALMA stacking of JADES galaxies}

We used the A$^3$GOODSS ALMA archival database, part of the A$^3$COSMOS database \citep[version 20220606;][]{Liu19,Adscheid24}, to perform a stacking analysis on our galaxy sample, choosing ALMA images from bands 6 ($\sim$243\,GHz) and 7 ($\sim$324\,GHz). 
Out of the 5974 galaxies in our $6<z<7$ sample, 4464 are covered by one or multiple ALMA pointings (i.e., within an area of a primary beam attenuation $>0.5$).
Regarding astrometry, the most critical optical versus ALMA astrometric errors are $<$0.2\arcsec \citep[e.g.,][]{Franco18}.
JWST astrometry are tied to Gaia and should therefore be better. 
As the ALMA observations we are using have PSF FWHM greater than 0.2\arcsec, we do not consider astrometry as a possible issue.
This reduces our sample to 4464 galaxies, with redshift, stellar mass, UV luminosity, and UV spectral slope distributions that are indistinguishable (Kolmogorov-Smirnov test) from those of their parent sample of 5974 galaxies.

Because of this large number of galaxies, it was impractical to stack them in the $uv$-domain in order to deal with the heterogeneity in angular resolution (ranging from $\sim$0.1$\arcsec$ to $\sim$2$\arcsec$) of these multiple ALMA pointings \citep[see, e.g.,][]{Wang22, Wang24,Magnelli24}. 
To solve this problem, we instead used the fact that at $\sim$1\,mm, galaxies are compact in size \citep[$\sim$0.1$\arcsec$; e.g.][]{GomezGuijarro22a} and can therefore be considered spatially unresolved at resolutions $\geq$0.2$\arcsec$, so that $S_{\rm peak}=S_{total}$.
Therefore, we excluded images with synthesised beam sizes $<0.2\arcsec$ from our analysis and `stacked' only primary beam-corrected flux densities measured from the central pixel at each source position. 
The measured fluxes were renormalised from their observed wavelength (within the ALMA bands 6 or 7) to a common 1.1\,mm observed wavelength using the star-forming galaxy SED template of \cite{Bethermin12}. 
The final stacked flux density is the noise-weighted mean of all individual flux densities, $S_i$: 
\begin{equation*}
S_{1.1\,mm,\mathrm{stack}} = \frac{\sum S_{1.1,i} \cdot \sigma_i^{-2} }{\sum\sigma_i^{-2}},
\end{equation*}
with $\sigma_i$ the pixel RMS noise at the position of each galaxy.

We first binned galaxies by UV magnitude in four bins: $-22<M_{\mathrm{UV}}<-20$~mag, $-20<M_{\mathrm{UV}}<-19$~mag, $-19<M_{\mathrm{UV}}<-18$~mag, and $-18<M_{\mathrm{UV}}<-17$~mag.
Selecting galaxies with $M_{\mathrm{UV}}<-17$~mag yields a total of 2455 galaxies.
We provide in Table~\ref{tab:stack_fluxes} the resulting ALMA measurements for each bin as well as the corresponding number of galaxies and pointings.

\begin{figure*}[!htbp] 
 	\includegraphics[width=\textwidth]{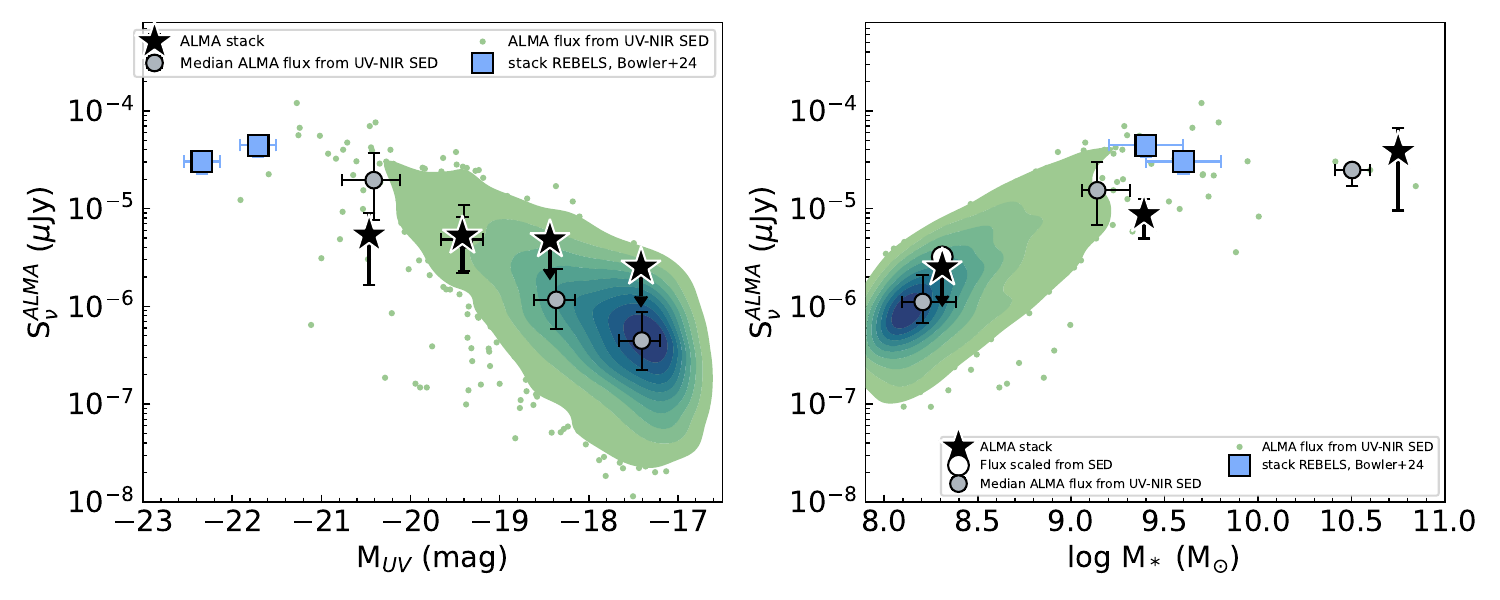}
  	\caption{\label{fig:flux} \textbf{Left panel:} Flux densities obtained by stacking ALMA observations in four UV magnitude bins (black symbols). Circles with downward arrows are 3-$\sigma$ upper limits. The contours show the spread of predicted fluxes without the ALMA constraint. The grey circles with black borders are the median values of these predicted fluxes in each UV magnitude bin. The blue square are the stack values of the REBELS sample \citep{Bowler24}. \textbf{Right panel:} Same as left panel but using three stellar mass bins. The white circle is the expected flux density assuming a simple scaling to the stellar mass of the $9<\log M_\star<10$ bin. }
\end{figure*}

\begin{table*}[!htbp]
    \centering
    \caption{Results of the ALMA stacking analysis at $6<z<7$.}
    \begin{tabular}{cccc}
    \hline
    \hline    
         &  N$_{galaxies}$&N$_{pointings}$ & $\left<S_{\nu}\right>$ \\
         &                  &               &$\mu$Jy\\
    \hline
    $-22<M_{\mathrm{UV}}<-20$     &   45& 198& 5.4$\pm$3.7\\
    $-20<M_{\mathrm{UV}}<-19$     &   154& 515& 5.2$\pm$3.0\\
    $-19<M_{\mathrm{UV}}<-18$     &   629& 2056& $<$4.8\\
    $-18<M_{\mathrm{UV}}<-17$     &   1627& 5943& $<$2.5\\
    \hline
    $8< \log$M$_\star<9$     &   1352& 5341& $<$2.5\\
    $9< \log$M$_\star<10$      &   59& 226& 8.7$\pm$3.8\\
    $10< \log$M$_\star<11$     &   4& 14& 38.5$\pm$29.0\\
    \hline
    \end{tabular}
    \label{tab:stack_fluxes}
\end{table*}
We show these measurements in the left panel of Fig.~\ref{fig:flux} along with the predicted ALMA fluxes obtained from UV-NIR only SED fitting.

The ALMA flux of the brightest sources ($M_{\mathrm{UV}}<-20$~mag) is slightly overestimated in the absence of IR constraint although still compatible within the errors.
At fainter magnitudes, the upper limit predictions are in agreement with the ALMA measurements.
We note that if we assume the ALMA fluxes predicted from the UV-NIR SED modeling, a S/N=3 detection from the ALMA stacking would require approximately 10660 and 52000 sources, in the $-19<M_{\mathrm{UV}}<-18$~mag and $-18<M_{\mathrm{UV}}<-17$~mag bins, respectively, or a deeper ALMA observations of the 629 and 1627 sources observed in these bins, respectively.

We also stacked the ALMA observations of the JADES/A$^3$COSMOS galaxies by bins of stellar mass ($8<\log M_\star/M_\odot<9$, $9<\log M_\star/M_\odot<10$, and $10<\log M_\star/M_\odot<11$), and show the results on the right panel of Fig.~\ref{fig:flux}.
Selecting galaxies with $\log M_\star/M_\odot>8$ yields a total number of 1415 galaxies.
The predictions from UV-NIR SED fitting are consistent with the constraints obtained from the ALMA stacking.
Despite being the most populated bin (1352 galaxies), the ALMA measurement of the lowest stellar mass bin ($8<\log M_\star/M_\odot<9$) results in a 3$\sigma$ upper limit of 2.5\,$\mu$Jy.
A factor 5 more sources would be needed according to SED modeling predictions to reach a S/N of 3 on this measurement.
These galaxies being less massive, it is expected that they contain less dust than more massive galaxies. 
We cannot constrain the dust mass of the galaxies of our sample since we only have one ALMA data point probing the 137-157\,$\mu$m rest-frame emission.
To understand if this non-detection is compatible with the fact that lower-mass galaxies should have lower dust content, we take the stellar mass as a proxy for galaxy SED normalisation and scale down the ALMA flux density obtained from galaxies in the intermediate mass bin ($9<\log M_\star/M_\odot<10$).
We obtain a predicted S$_{1.1\,mm}$ of 3.2\,$\mu$Jy which is tension with our limit, suggesting that the dust content per unit stellar mass of these low mass systems is indeed lower than that observed in more massive objects.

\section{\label{impact}Impact of the ALMA constraint on physical properties of \texorpdfstring{$6<z<7$}{6<z<7} galaxies}

In this section, we investigate the impact of the ALMA constraints on the inference of the physical properties of the JADES subsample.
We thus run \texttt{CIGALE} a second time using the ALMA stacking measurements as constraints for the IR part of the SED.
We attribute to each galaxy the ALMA flux density obtained from stacking according to the derived UV magnitude of the first \texttt{CIGALE} run.
For example, each galaxy with a UV magnitude within $-22<M_{\mathrm{UV}}<-20$ is attributed a S$_{1.1\,mm}$ of 5.4$\pm$3.7\,$\mu$Jy.

\begin{figure}[!htbp] 
 	\includegraphics[width=\columnwidth]{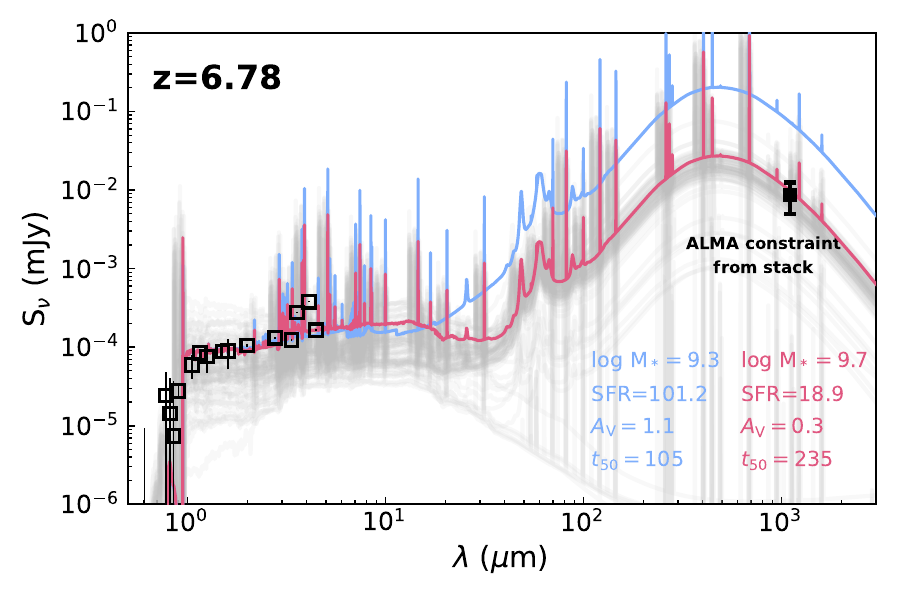}
  	\caption{\label{fig:sed} Example of SED for a source at $z=6.78$ in the $9<\log (M_\star^{\mathrm{no-IR}}/M_\odot)<10$ stellar mass bin. The black open squares are the fluxes from HST and JWST and the blue line is the best fit model obtained fitting these data points. The filled black square indicated the ALMA measurement from stacks obtained in the same stellar mass bin, and the red line indicate the best fit model obtained when adding this IR constraint to the SED fitting. Parameters obtained from the two fits are indicated using the same colours. As an indication, the best SEDs obtained from the rest of the galaxies in the same stellar mass bin are shown in light grey.}
\end{figure}

In Fig.~\ref{fig:sed}, we show an example of a $z=6.78$ galaxy with $M_{\mathrm{UV}}=-20.5$\,mag and $\log M_\star^{\mathrm{no-IR}}=$9.3.
We show on the figure the resulted best fits without and with the ALMA constraint in blue and red, respectively.
Clearly, without the ALMA constraint, the IR luminosity is over-predicted by the UV-NIR SED modelling, resulting in a significant A$_{\rm V}$ of 1.1\,mag.
Using ALMA stacked measurement as a constraint, the A$_{\rm V}$ decreases to 0.3\,mag.
This difference in A$_{\rm V}$ results in a difference in age, with an older age obtained from the run using ALMA of 235\,Myr compared to 105\,Myr when ALMA is not used.

\begin{figure*}[!htbp] 
 	\includegraphics[width=\textwidth]{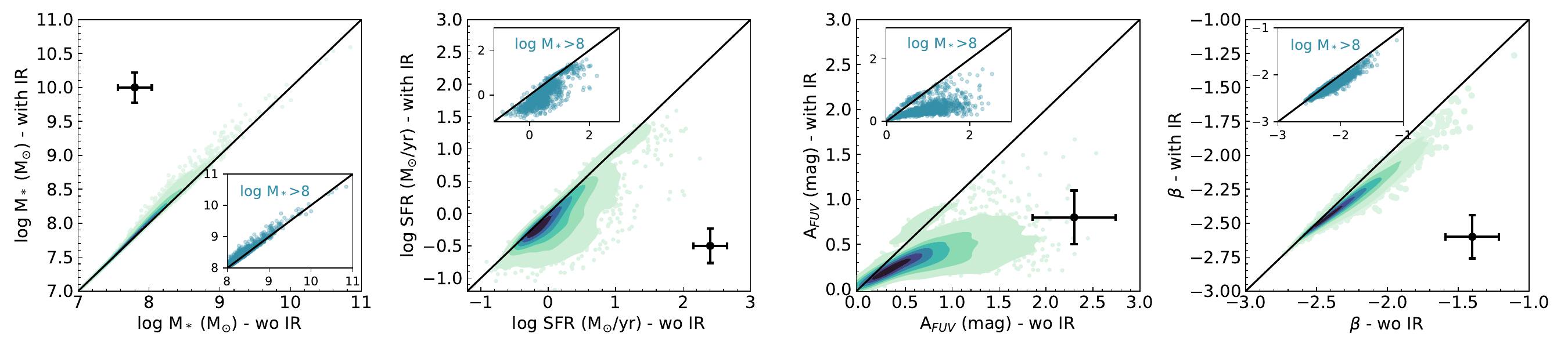}
  	\caption{\label{fig:comps} Comparison between the physical properties obtained when fitting the UV-to-NIR restframe observations only (HST+JWST) and the ones obtained when adding the ALMA constraints. The black solid line is the one-to-one relationship. Results for the stellar mass, SFR, FUV attenuation, and $\beta$ are shown from left to right. The black circles with errorbars indicate the median error in each panel. The inset panels show the same figure for the most massive ($\log M_\star^{noIR}/M_\odot>8$) sources of the sample.}
\end{figure*}

To understand the impact of the IR constraint on the derived physical properties of the galaxies of the whole sample, we compare in Fig.~\ref{fig:comps} the stellar masses, SFRs, FUV attenuations, and UV continuum $\beta$ slopes obtained with and without the ALMA constraint.

The stellar mass is weakly affected by the addition of the ALMA constraints.
However, for galaxies with $8<\log (M_\star^{\mathrm{no-IR}}/M_\odot)<10$ the use of ALMA leads to slightly higher stellar masses, although it is well within the typical error on $M_\star$ which is 0.24\,dex.
Regarding the SFR, the estimates with and without IR measurements are on average in agreement although there are sources, the most massive ones ($\log (M_\star^{\mathrm{no-IR}}/M_\odot)>8$ for which the SFR is clearly lower using IR measurements, with a median difference of -0.37\,dex.
For the FUV attenuation, the average trend shows a lower $A_{\rm FUV}$ when ALMA is used but only by 5\% on the average value, which is negligible compared to the typical errors on this measurement.
However, as for the SFR, we find that for the most massive sources, $A_{\rm FUV}$ is also lower when using IR constraints, with a median value of -0.49\,mag. 
Finally, the UV slope $\beta$ tends to be bluer when using ALMA constraints.
The sources that deviate the most from the one-to-one relationship for each parameter are galaxies with $\log (M_\star^{noIR}/M_\odot)>8$.
For these sources, without IR constraint, the code is using stronger attenuation to fit the observations while the fit with IR tends toward lower attenuation but older ages instead.
This comparison suggests the evidence of biases when estimating the physical properties of $z>6$ galaxies without FIR constraints.

\section{\label{dust_results}Dust attenuation properties}

In this section, we study the dust attenuation properties of $6<z<7$ galaxies, taking into account the ALMA constraint obtained from stacking.
The physical properties of this sample are derived from their distribution (median, 25$^{th}$, and 75$^{th}$ percentiles) using the same weights used to estimate the ALMA fluxes and are provided in Table~\ref{tab:stack_prop}.
For bins containing both galaxies with high S/N values and galaxies with upper limits on these physical properties, we derive upper limits of the median value using a parametric approach, that is by fitting a normal distribution to the data and determining the z-score at 84$\%$ confidence level. The upper limit of the bin is then derived by adding to the mean of the distribution the z-score times the standard deviation of the distribution.

\begin{table*}[!htbp]
    \centering
    \caption{Stacked physical properties at $6<z<7$, derived from the SED modelling that includes the ALMA constraint.}
    \begin{tabular}{cccccccc}
    \hline
    \hline    
         &  $\left<\textrm{redshift}\right>$&$\log\left<M_\star\right>$& $\left<\textrm{SFR}\right>$& $\left<\beta\right>$ & $\left<M_{\mathrm{UV}}\right>$& $\left<\textrm{IRX}\right>$ &log $\left<L_{\mathrm{IR}}\right>$\\
         &  &M$_{\odot}$          & M$_{\odot}$yr$^{-1}$    &           & mag       &    &L$_{\odot}$\\
    \hline
    $-22<M_{\mathrm{UV}}<-20$     &  $6.18^{+0.51}_{-0.06}$&$9.43^{+0.0}_{-0.14}$ & $14.6^{+1.3}_{-6.7}$& $-2.05^{+0.01}_{-0.03}$& $-20.47^{+0.38}_{-0.07}$& $0.05^{+0.12}_{-0.09}$& $10.53^{+0.06}_{-0.06}$\\
    $-20<M_{\mathrm{UV}}<-19$     &  $6.64^{+0.09}_{-0.42}$&$8.85^{+0.12}_{-0.51}$& $12.6^{+6.8}_{-4.6}$& $-2.13^{+0.05}_{-0.18}$& $-19.44^{+0.19}_{-0.25}$& $0.30^{+0.16}_{-0.15}$& $10.42^{+0.25}_{-0.18}$\\
    $-19<M_{\mathrm{UV}}<-18$     &  $6.68^{+0.20}_{-0.38}$&$8.23^{+0.31}_{-0.07}$& $2.0^{+0.7}_{-1.1}$& $-2.34^{+0.07}_{-0.07}$& $-18.40^{+0.25}_{-0.25}$& $<$-0.09& $<$9.59\\
    $-18<M_{\mathrm{UV}}<-17$     &  $6.62^{+0.12}_{-0.36}$&$7.99^{+0.21}_{-0.22}$& $0.7^{+0.3}_{-0.2}$& $-2.28^{+0.10}_{-0.12}$& $-17.38^{+0.19}_{-0.28}$& $<$-0.10&$<$9.17\\
    \hline
    $8<\log M_\star<9$     &   $6.65^{+0.14}_{-0.36}$&$8.27^{+0.15}_{-0.12}$& $0.4^{+0.4}_{-0.2}$& $-2.16^{+0.22}_{-0.14}$& $-17.47^{+0.59}_{-0.69}$&  $<$0.11&$<$9.20\\
    $9<\log M_\star<10$     &   $6.19^{+0.53}_{-0.07}$&$9.27^{+0.14}_{-0.19}$& $8.8^{+8.3}_{-7.2}$& $-2.03^{+0.08}_{-0.13}$& $-20.06^{+0.43}_{-0.48}$&  $0.13^{+0.05}_{-1.82}$& $10.59^{+0.12}_{-2.30}$\\
    $10<\log M_\star<11$     &   $6.31^{+0.00}_{-0.00}$&$10.86^{+0.00}_{-0.44}$& $9.2^{+0.0}_{-0.0}$& $-0.42^{+0.00}_{-0.67}$& $-18.37^{+0.00}_{-0.96}$&  $1.28^{+0.00}_{-0.24}$& $10.96^{+0.15}_{-0.00}$\\
    \hline
    \end{tabular}
    \label{tab:stack_prop}
\end{table*}

\begin{figure}[!htbp] 
 	\includegraphics[width=\columnwidth]{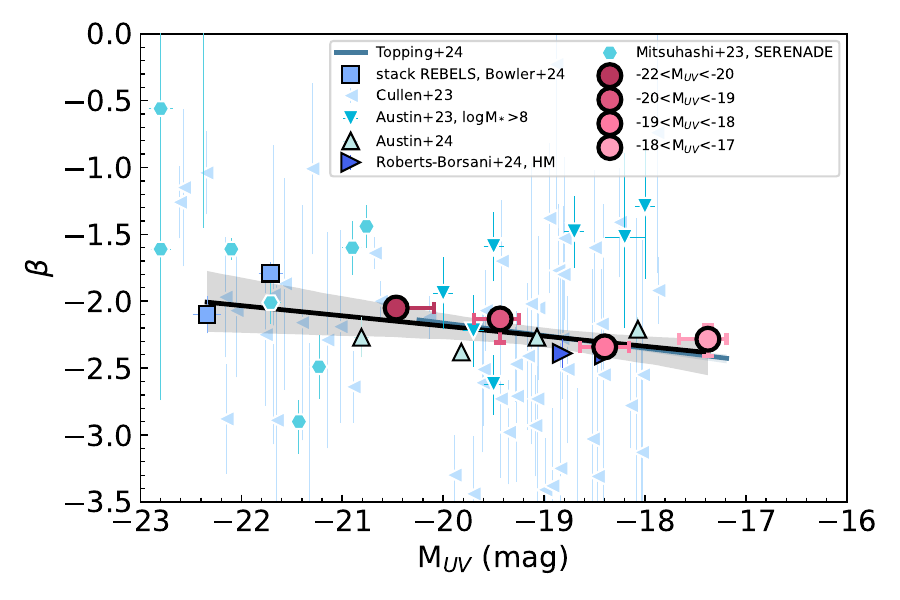}
  	\caption{\label{fig:beta_muv} $\beta$ as a function of UV magnitude. Big circles with black edges are weighted medians in 4 four UV magnitude bins. Samples from the literature are indicated by symbols with different shades of blue \citep{Bowler24,Cullen23,Austin23,RobertsBorsani24,Mitsuhashi24}, white edges showing individual galaxies while black edges are stacked or median values. The relation from \cite{Topping24} is shown with the solid line. }
\end{figure}

Several pre-JWST and JWST studies have reported a relation between the UV $\beta$ slope and UV magnitude such as UV slopes shift towards bluer $\beta$ at low UV luminosities \citep[e.g.,][]{Finkelstein12,Bouwens14,BhatawdekarConselice21,Topping24}.
Although still debated at $z>8$, this relation would indicate a trend between dust attenuation, which is the dominant component affecting the UV slope, and UV luminosity that may hint at a relationship between luminosity and metals in early galaxies \citep{Cullen23,Topping24}. 
In Fig.~\ref{fig:beta_muv}, we compare the UV properties ($\beta$ and $M_{\mathrm{UV}}$) of our sample with sources from the literature at the same redshift.

Combining our measurement to the stack estimates from REBELS \citep{Bowler24}, \cite{Austin24}, and \cite{RobertsBorsani24}, we obtain the following relation from a linear regression fit:
\begin{equation}
\beta = -0.08\pm0.03 \times M_{\mathrm{UV}} -3.70\pm0.11.
\end{equation}
The resulting slope is in agreement with the relations from \cite{Topping24} who found a $d\beta/dM_{\rm UV}$ slope of $-$0.11$\pm$0.02 and $-$0.12$\pm$0.02 at $z\sim5.9$ and $z\sim7.3$, respectively, as well as the slopes obtained at $z\sim5$ by \cite{Rogers14} and \cite{Bouwens14} within errors (-0.12$\pm$0.02 and -0.14$\pm$0.02, respectively).

\begin{figure}[!htbp] 
 	\includegraphics[width=\columnwidth]{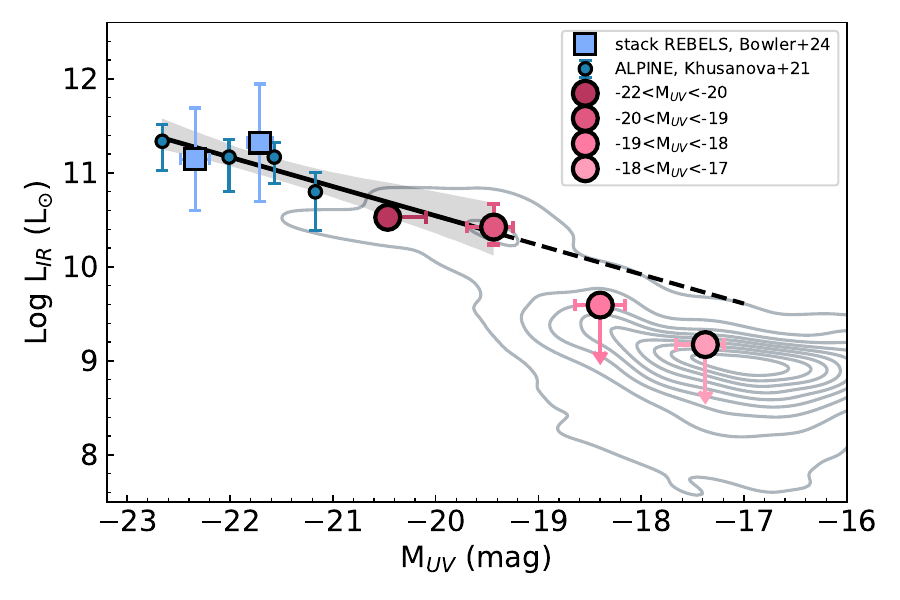}
  	\caption{\label{fig:lir_muv} Infrared luminosity as a function of UV magnitude. The contours show the position of the whole sample with $L_{\mathrm{IR}}$ obtained from the fit using the ALMA constraint. Big circles with black borders are the weighted median values in four UV magnitude bins. Stacked or median values for samples from the literature are show in various shades of blue symbols \citep{Bowler24,Khusanova21}. }
\end{figure}

The ALMA stacking provides a constraint on the IR luminosity of the galaxies of our sample.
In Fig.~\ref{fig:lir_muv}, we show the $L_{\mathrm{IR}}$ as a function of UV magnitude.
The JADES sample combined with A$^3$COSMOS observations allows us to probe fainter magnitudes and extend the $L_{\mathrm{IR}}$ vs M$_{UV}$ relation down to M$_{UV}\sim-17$~mag.
Our sample has $\log L_{\mathrm{IR}}/L_\odot$ spanning from 7.4 to 12.3 with a median value of 9.3.
For comparison, we add the stacked measurements of the REBELS sample as well as those from the ALPINE sample at $z\sim5.5$ from \cite{Khusanova21}.
Both these samples probe higher UV magnitude ranges and higher IR luminosities.
A fit to the REBELS+ALPINE stacked signals combined with our two highest UV magnitude bins yields the following relation: 

\begin{equation}
\log \left(L_{\mathrm{IR}}/L_{\odot}\right) = -0.31\pm0.05 \times M_{\mathrm{UV}} + 4.31\pm0.42.
\end{equation}

Extended to fainter UV magnitude, this relation predicts higher $L_{\mathrm{IR}}$ at M$_{UV}>$-19 than our estimates using ALMA.
At M$_{UV}=$-18.4~mag and -17.4~mag, the relation derived from higher UV magnitudes overestimates the $L_{\mathrm{IR}}$ by 0.45\,dex and 0.56\,dex, respectively. 
This is a possible hint at a breakdown in the $L_{\mathrm{IR}}$ vs M$_{UV}$ relation at low UV magnitudes that could be due to lower dust content. 
For the two UV faintest bin that have a median M$_{UV}$ of -18.4 and -17,4, the average stellar mass are $\log M_{\star}/M_\odot = 8.2$ and 8.0, respectively. 
These two UV bins have also low dust attenuation: 0.31 and 0.33, respectively.
Therefore, at these low UV magnitude, we are dominated by low-mass, low-attenuation systems, leading to a turnover of L$_{IR}$, which has been already observed at lower redshift ($z\sim4$) by \cite{Bernhard14}.

\subsection{IRX versus \texorpdfstring{$\beta$}{beta}}

\begin{figure}[!htbp] 
 	\includegraphics[width=\columnwidth]{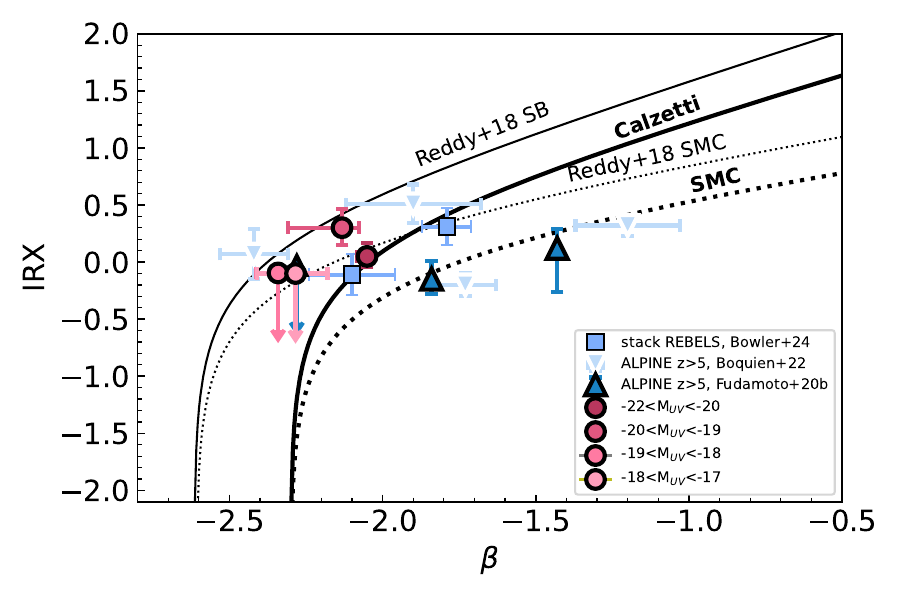}
  	\caption{\label{fig:irx_beta} IRX as a function of $\beta$. Coloured circles with black borders are weighted medians derived in four UV magnitude bins from our sample.  The blue squares are the stacked values of \cite{Bowler24} for the REBELS sample. The light blue downward triangles are the ALPINE $z>5$ from \cite{Boquien22} while the dark blue triangles are ALPINE stacks at $z>5$ values from \cite{Fudamoto20}. The thick black solid line is the relation assuming a \cite{Calzetti00} attenuation law while the dotted line is assuming a SMC-like extinction curve. The  thin black lines are the corresponding relations from \cite{Reddy18}.}
\end{figure}

We place our measurements, obtained from ALMA stacking in UV magnitude bins, on the IRX-$\beta$ diagram in Fig.~\ref{fig:irx_beta}.
Our averaged data points are close to the stacked measurements from REBELS, especially the brightest UV magnitude bin ($-22<M_{\mathrm{UV}}<-20$) which corresponds to the UV magnitude range probed by the REBELS galaxies.
They disfavour the expected relation from a SMC extinction curve but are compatible with a \cite{Calzetti00} relations used in \cite{Bowler24} with $\beta_0=-2.3$.
Although our SFH library has a wide range of $\beta_0$, galaxies of our sample favour blue $\beta_0$ with a median value of -2.5, which is consistent with the assumed $\beta_0$ of \cite{Reddy18}.
Therefore, including the effect of dust, our sample has relatively blue $\beta$ with a median value of -2.25.

Galaxies of our sample are also compatible with both the starburst and SMC relations of \cite{Reddy18} who used them with a different $\beta_0$ than \cite{Bowler24} ($\beta_0=-2.6$).
These relations were found by \cite{Boquien22} to be suitable for the ALPINE sample.
In our process, we tested letting the $\delta$ slope of the attenuation curve free, like in \cite{Boquien22}.
However, the improvement of the fit, using a $\Delta$BIC analysis (Bayesian Information Criterion), proved to be statistically significant for only 3.7$\%$ of the sample.
Therefore, we kept $\delta$ fixed to 0 (no modification of the attenuation curve).
Although they seem to favour the \cite{Reddy18} relations, most of estimates at low IRX are obtained from the ALMA upper limits on the IR and are thus upper limits on the IRX, as indicated by the two faintest UV magnitude bins.
However, at higher IRX, our estimates are compatible with both a \cite{Calzetti00} with $\beta_0=-2.3$ and \cite{Reddy18} relations (starbursts and SMC, using $\beta_0=-2.6$).
These findings are consistent with the results that grain size and dust composition at very high redshift are similar to local starbursts and SMC, indicating no fundamental evolution.
Our results confirm that at $z>6$, there is no clear sign of a break-down of \cite{Calzetti00} relation.

\subsection{IRX versus stellar mass}

\begin{figure}[!htbp]
 	\includegraphics[width=\columnwidth]{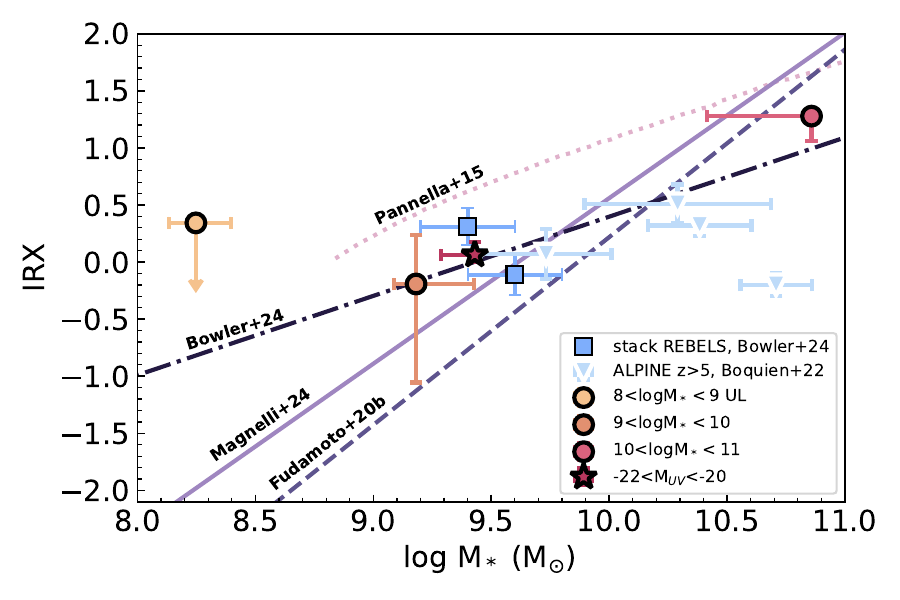}
  	\caption{\label{fig:irx_mstar} IRX as a function of stellar mass. Circles with black edges are weighted medians colour-coded according to the stellar mass bin used for stacking. The red star is the position of the $-22<M_{UV}<-20$ bin for direct comparison with the stacked values of \cite{Bowler24} for the REBELS sample (blue squares). The light blue downward triangles are the ALPINE $z>5$ stacks values from \cite{Boquien22}. The solid light purple line is the relation from \cite{Magnelli24}, the dashed line is from \cite{Fudamoto20}, the dashed-dotted line is the relation from \cite{Bowler24}, and the dotted light purple line the relation from \cite{Pannella15} at lower redshifts. The darker the color of these relations, the higher the redshifts of galaxies used to derive them.}
\end{figure}

We place our galaxies on the IRX-$M_\star$ plane in Fig.~\ref{fig:irx_mstar}. 
Specifically in this figure, we use the results of the SED fitting using the ALMA constraints obtained from stacking in bins of stellar mass ($8<\log M_\star/M_\odot<9$, $9<\log M_\star/M_\odot<10$, and $10<\log M_\star/M_\odot<11$).
Along with our data points, we add the relations from \cite{Pannella15} at $z\sim 1-3$, \cite{Magnelli24} at $4<z<5$, \cite{Fudamoto20} at $z\sim 5.5$, and \cite{Bowler24} at $z\sim 7$.

Our low mass bin upper limit on IRX does not provide any useful constraint on the position of these galaxies on the IRX-$M_\star$ plane.
The intermediate mass median value is compatible with the relations of \cite{Bowler24}, \cite{Magnelli24}, and marginally with the relation of \cite{Fudamoto20}.
There seems to be a better agreement with the relation of \cite{Bowler24} albeit the large scatter indicated by the errorbars (25$^{th}$ and 75$^{th}$ percentile).
The high mass bin data point is in agreement with both \cite{Magnelli24} and \cite{Fudamoto20} relations, and marginally with the \cite{Bowler24} relation, but here again, the large scatter does not allow us to discriminate a relation.
For comparison, we show the brightest UV magnitude bin, $-22<M_{UV}<-20$ (red star), for consistency with the REBELS sample and find that it falls on the relation from \cite{Bowler24}, thus in agreement with their result.
Finally, we show the relation obtained by \cite{Pannella15} at lower redshift ($z<4$), indicating that there is a clear decrease of IRX at fixed stellar mass with increasing redshift.

\subsection{\label{tdust}Impact of dust temperature assumption}

At $6<z<7$, ALMA 1.1\,mm probes the 137-157\,$\mu$m rest frame which provides a good constraint on the peak of the IR SED. 
However, with only one flux measurement in the IR, the $L_{\mathrm{IR}}$ strongly depends on the assumption taken on the IR SED shape. 
In the dust emission model of \cite{Schreiber18}, this shape is handled through two free parameters: the PAH fraction and the dust temperature.
The PAH fraction has little effect on the $L_{\mathrm{IR}}$, but the dust temperature, that we originally fixed at 47\,K, is a key assumption that we can test in this section.
We run \texttt{CIGALE} several times, each time fixing the dust temperature to a different value.
Given their extreme $L_{\mathrm{IR}}$ (Fig.~\ref{fig:lir_muv}), on average higher than the galaxies of our sample, we consider the dust temperature of REBELS galaxies determined by \cite{Sommovigo22} to be the maximum dust temperature for our galaxies.
We thus only decrease T$_{\mathrm{dust}}$ from the original assumption (47\,K) down to a minimum value of 30\,K, which corresponds to the lowest temperature of REBELS galaxies found by \cite{Sommovigo22} (REBELS-32, 39\,K minus the lower error of 9\,K).
We note that \cite{Mitsuhashi24} found temperatures as low as 26$^{+14.5}_{-6.5}$\,K in the SERENADE sample, which is typical of dust temperature at $z<1$ \cite[e.g.,][]{Schreiber18}, and, recently, \cite{Algera24} obtained a dust temperature of 32$^{+9}_{-6}$\,K for REBELS-25, a galaxy at redshift 7.31.
Our lowest assumed temperature of 30\,K is thus also consistent with the temperatures derived from the SERENADE sample and REBELS-25.

\begin{figure}[!htbp] 
 	\includegraphics[width=\columnwidth]{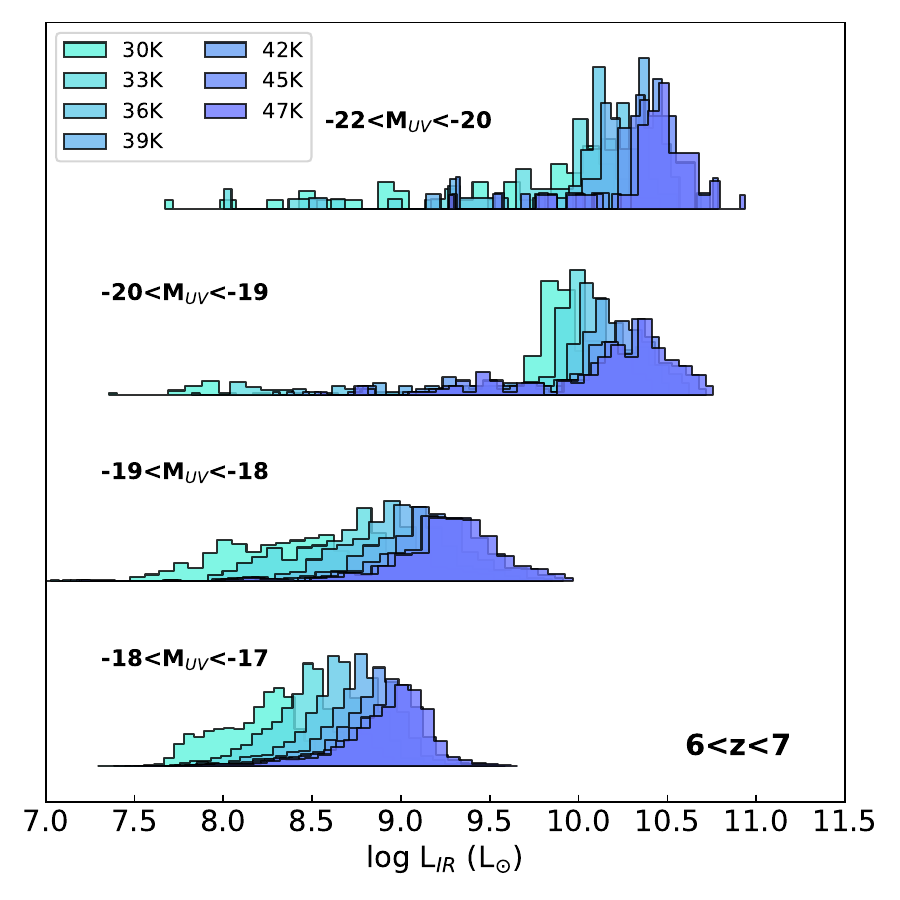}
  	\caption{\label{fig:lir_muv_lowT} $L_{\mathrm{IR}}$ distributions of galaxies in the four UV magnitude bins considered in this work, assuming different dust temperatures.}
\end{figure}

\begin{figure}[!htbp] 
 	\includegraphics[width=\columnwidth]{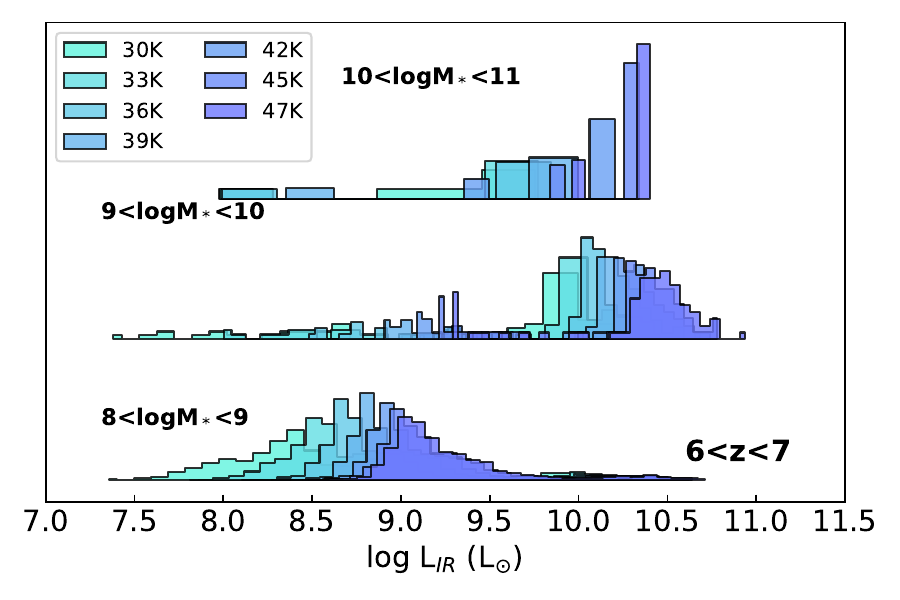}
  	\caption{\label{fig:lir} $L_{\mathrm{IR}}$ distributions of galaxies in the three stellar mass bins considered in this work, assuming different dust temperatures. }
\end{figure}

We look at the distribution of $L_{\mathrm{IR}}$ using either the UV magnitude bins for the ALMA stacking (Fig.~\ref{fig:lir_muv_lowT}) or stellar mass bins (Fig.~\ref{fig:lir}).
We show the results of the SED modelling run for each dust temperature assumption (30, 33, 36, 39, 42, 45, and 47\,K).
As expected, the effect of decreasing dust temperature assumption is to shift the $L_{\mathrm{IR}}$ distributions towards lower values. 
However, temperatures higher than $\sim$36\,K result in similar $L_{\mathrm{IR}}$.
This is due to the fact that the rest frame range probed by the ALMA constraint at these redshifts is close to the peak of the IR SED at high temperature.
Therefore, if the dust temperature in these galaxies is higher than $\sim$36\,K (value used in our test), then the $L_{\mathrm{IR}}$ estimate is quite reliable.
Despite these uncertainties on the IR SED, the distributions shown in Fig.~\ref{fig:lir_muv_lowT} and Fig.~\ref{fig:lir} provide average constraints on the range of $L_{\mathrm{IR}}$ of the bulk of galaxies at $6<z<7$.

\begin{figure*}[!htbp] 
 	\includegraphics[width=\textwidth]{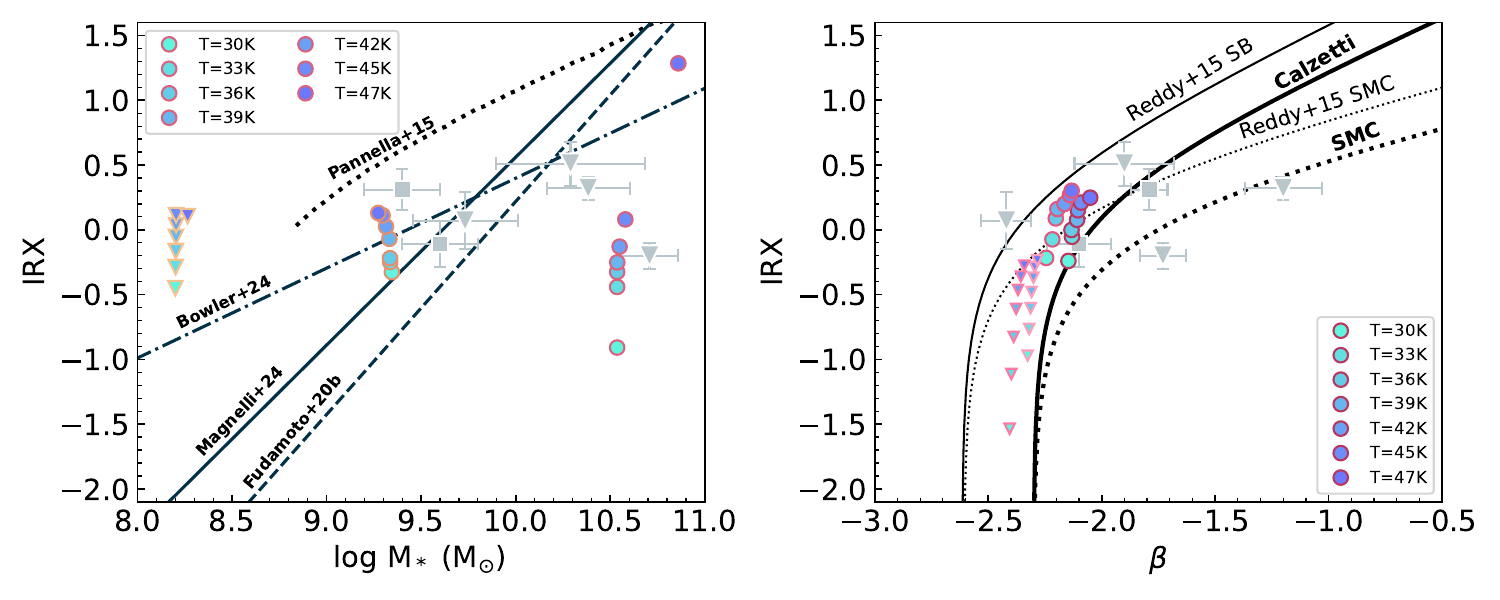}
  	\caption{\label{fig:irx_mstar_beta_lowT} \textbf{Left panel:} IRX as a function of stellar mass (symbols and lines following Fig.~\ref{fig:irx_mstar}). Filled colour circles are the stacked measurements obtained for each temperature assumption. \textbf{Right panel:} IRX-$\beta$ diagram (symbols and lines following Fig.~\ref{fig:irx_beta}). Galaxies are colour-coded as in the left panel.}
\end{figure*}

The impact of T$_{\mathrm{dust}}$ on the different physical properties studied in Sect.~\ref{dust_results} is presented in Fig.~\ref{fig:irx_mstar_beta_lowT}.
This figure shows on the left panel the IRX-M$_\star$ plane using ALMA observations stacked in stellar mass bins, assuming the same set of dust temperatures than in Fig.~\ref{fig:lir_muv_lowT}.
As expected, the IRX parameter is affected. 
Very small variations are also found in M$_\star$ due to the fact that \texttt{CIGALE} performs SED modelling using energy balance over the whole wavelength range, another indication on the error on the measurement that we can have on this parameter.
Despite the wide range of dust temperature, the intermediate mass bin galaxies are still compatible with the \cite{Bowler24} relation.
A strong effect is however observed for the most massive bin, but this bin contains only 4 galaxies and is thus more subject to statistical variations.

On the right panel of Fig.~\ref{fig:irx_mstar_beta_lowT}, we show the impact of dust temperature assumption on the IRX-$\beta$ diagram.
The conclusions of Sect.~\ref{dust_results} are still valid: 
Our data points are compatible with a \cite{Calzetti00} relation using $\beta_0=-2.3$ or the \cite{Reddy18} relation for starburst and SMC with $\beta_0=-2.6$.
Even with a more accurate estimate on the dust temperature, it will be difficult to discriminate, within the errors, between these relations in this $\beta$ range.

\section{\label{highz}Infrared constraints at \texorpdfstring{$z>7$}{z>7}}

We apply the ALMA stacking analysis described in Sect.~\ref{stack} on JADES galaxies at $z>7$ . 
The measurements result only in upper limits, however they provide first constraints on the IR emission of the bulk of the population at these redshifts.
We provide these upper limits in Table~\ref{tab:stack_fluxes_z7} in UV magnitude and stellar mass bins.
We then perform the same SED modelling procedure on the $z>7$ sources as for the $6<z<7$ redshift bin, assuming the same dust temperature of $T_{\rm dust}=$47\,K.

\begin{table}[!htbp]
    \centering
    \caption{Results of the ALMA stacking analysis at $7<z<9$ and $9<z<12$.}
    \begin{tabular}{cccc}
    \hline 
         &   N$_{galaxies}$&N$_{pointings}$ & $\left<S_{\nu}\right>$ \\
 & &  &$\mu$Jy\\
    \hline 
    \hline
     \multicolumn{4}{c}{\textbf{7$<$z$<$9}}\\
    \hline   
     $-22<M_{\mathrm{UV}}<-20$ & 17 & 55 &$<$26.4\\
     $-20<M_{\mathrm{UV}}<-19$& 90& 267 &$<$13.6\\
     $-19<M_{\mathrm{UV}}<-18$& 302& 1058 &$<$6.2\\
     $-18<M_{\mathrm{UV}}<-17$& 893& 3337 &$<$3.2\\
    \hline
    $8<$ log M$_\star<9$     &    863&3527 &$<$2.9\\
    $9<$ log M$_\star<10$     &    35&130 &$<$14.8\\
    $10<$ log M$_\star<11$     &    4&18 &$<$72.0\\
    \hline  
    \hline
     \multicolumn{4}{c}{\textbf{9$<$z$<$12}}\\
    \hline    
     $-22<M_{\mathrm{UV}}<-20$ & 7 & 18 &$<$62.3\\
     $-20<M_{\mathrm{UV}}<-19$& 40 & 170 &$<$11.0\\
     $-19<M_{\mathrm{UV}}<-18$& 159& 479 &$<$9.4\\
     $-18<M_{\mathrm{UV}}<-17$& 403& 1447 &$<$4.5\\
    \hline
    $8<$ log M$_\star<9$     &    452&1734 &$<$3.8\\
    $9<$ log M$_\star<10$     &    33&127 & $<$14.4\\
    $10<$ log M$_\star<11$     &    2&4 &$<$188.0\\
    \hline
    \end{tabular}
    \label{tab:stack_fluxes_z7}
\end{table}

\subsection{Evolution of $\beta$}

\begin{figure}[!htbp] 
 	\includegraphics[width=\columnwidth]{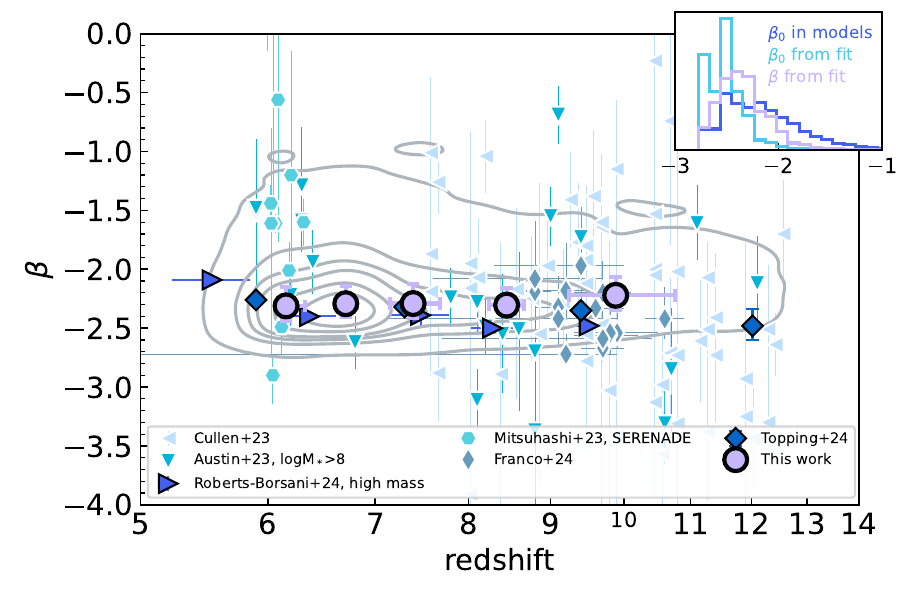}
  	\caption{\label{fig:betaz} $\beta$ as a function of redshift. Grey contours show the full JADES/A$^3$COSMOS sample while the purple circles show the median position in several redshift bins. Different samples from the literature are shown in shades of blue \citep{Laporte17,Laporte19,Tamura19,Inami22,Mitsuhashi24,Schouws24,Topping24}. The inset panel shows the distribution of the intrinsic UV slope, $\beta_0$, allowed by the models (dark blue), as well as $\beta_0$ and $\beta$ obtained from the SED fitting.}
\end{figure}

We examine $\beta$ as a function of redshift in Fig.~\ref{fig:betaz}.
Although galaxies of our sample are slightly redder than the averaged measurements of \cite{RobertsBorsani24} at the same redshifts, they do not show extreme values and lie within the range spanned by individual galaxies in the literature.
As discussed in \cite{Franco24} \citep[see also][]{Topping24}, there is a caveat in using SED modelling to measure $\beta$ as models are limited to a given $\beta$ value toward the blue side of the distribution, in our case this value is defined by the lowest intrinsic value allowed by our model: $\beta_0=-2.77$.
On the other hand, despite being more flexible, a direct measurement of $\beta$ from the data can also introduce systematics due to the different rest-frame wavelength range used to measure $\beta$.

To understand to which extent our results could be affected by the lower limit on $\beta_0$ induced by SED modelling, we show in the inset panel of Fig.~\ref{fig:betaz} the distributions of $\beta_0$ allowed by our models (dark blue), $\beta_0$ obtained from the SED fitting run (cyan), the $\beta$ for our galaxies (purple).
For the total sample, that is the three redshift bins, only 15$\%$ of the galaxies select the lowest possible value of $\beta_0$.
A Kolmogorov-Smirnov test indicate that these galaxies have similar redshift and $A_{\rm FUV}$ distributions to the total sample, but different $M_\star$ and ages distributions resulting in lower values for both properties.
The fact that galaxies with the lowest $\beta_0$ are not at a particular redshift means that these galaxies would not necessarily drive a trend that we would be missing between $\beta$ and redshift.
Furthermore, although we would expect to miss galaxies with the bluest $\beta$ in the highest redshift bin from the expected trend, only 5$\%$ of these $z=$9-12 galaxies have the lowest $\beta_0$ value compared to 18 and 13$\%$ at $z=$6-7 and $z=$7-9, respectively.
Therefore, despite the fact that we may miss a fraction of galaxies with extreme blue colours, our results seem to favour a relatively flat relation between $\beta$ and redshift.

The $\beta$ distribution of our sample points toward an absence of trend between $\beta$ and redshift. 
A trend would be expected with galaxies having bluer colours with increasing redshift since they become progressively metal- and dust-poor \citep[e.g.,][]{Topping24,Cullen23}.
Recently, \cite{Cullen23} found extremely blue galaxies at $z\sim$12 with a median $\beta$ of -2.5 while \cite{Topping24} obtained a median $\beta$ of -2.48$^{+0.14}_{-0.12}$ on a sample drawn from JADES.
At the same redshift, we find a median value of -2.14$\pm$0.21 which is slightly redder.
Although both the studies from \cite{Cullen23} and \cite{Topping24} found a decrease of $\beta$ with increasing redshift, the latter mention that the rate of this evolution is slowest at low UV luminosities with minor differences between galaxies with $M_{UV}=-18$ at $z=$5-9 and $z>9$.
The bulk of our sample are UV faint (see Fig.~\ref{fig:lir_muv}) and the absence of strong evolution that we observe for the whole sample is thus consistent with the finding from \cite{Topping24}.
Interestingly, we note that $\beta$ is slightly redder in the highest redshift bin ($z>9.5$) possibly confirming the results of \cite{Saxena24} who found such a redder $\beta$ at the same redshift using a spectroscopic sample.
They interpret this redder value from a rapid build-up of dust content in the very early Universe or a significant
contribution from the nebular continuum emission to the observed UV spectra, depending on gas temperatures and densities. 

\begin{figure}[!htbp] 
 	\includegraphics[width=\columnwidth]{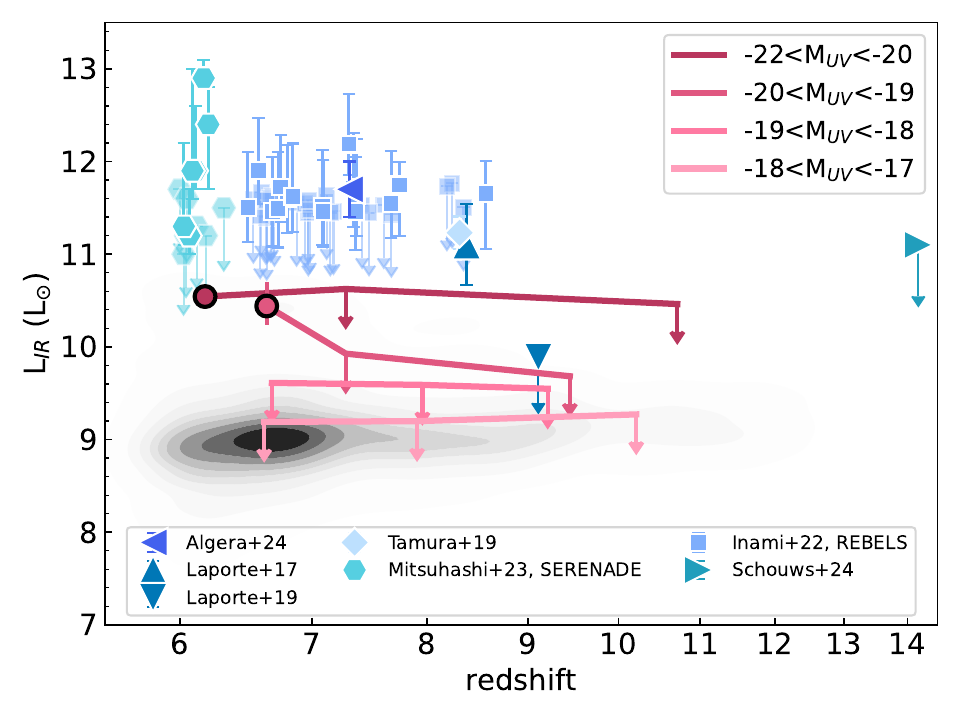}
  	\caption{\label{fig:lirz} IR luminosity as a function of redshift for our sample (black contours, downward arrows indicate upper limits) and different samples from the literature \citep[shades of blue,][]{Laporte17,Laporte19,Tamura19,Inami22,Mitsuhashi24,Schouws24}. Symbols with downward arrows are upper limits. Circles with black border are our fiducial UV magnitude bins.}
\end{figure}

We now try to put constraints on the evolution of $L_{\mathrm{IR}}$ with redshift.
At these high redshifts, the temperature of the Cosmic Microwave Background (CMB) increases rapidly reaching 19\,K, 22\,K, 27\,K, and 30\,K at $z=$6, 7, 9, and 10, respectively \citep{DaCunha13}.
If not taken into account in the SED fitting procedure, the impact of the CMB results in an under-estimation of the $L_{\mathrm{IR}}$ that depends on the dust temperature and redshift.
\cite{DaCunha13} demonstrated that for $T_{\rm dust}=40$\,K, the $L_{\mathrm{IR}}$ is under-estimated by -0.018, -0.019, -0.045, and -0.070\,dex at $z=$6, 7, 9, and 10, respectively.
We thus correct our $L_{\mathrm{IR}}$ measurement accordingly in Fig.~\ref{fig:lirz} where we show them as a function of redshift.
We note that at $6<z<7$ the correction is negligible and was not taken into account in Fig.~\ref{fig:lir_muv}.
For comparison, we add on Fig.~\ref{fig:lirz} individual measurements from the literature: the REBELS sample \citep{Inami22}, the SERENADE sample \citep{Mitsuhashi24}, and the individual sources of \cite{Laporte17}, \cite{Tamura19}, \cite{Laporte19}, and \cite{Schouws24}.
Fig.~\ref{fig:lirz} highlights the fact that the bulk of the population is still out of reach and that the samples on which models are basing their assumptions have extreme IR properties.

\subsection{Dust formation in early galaxies}

Combining V-band attenuation and sSFR measured from JWST/NIRSpec spectra of 631 galaxies at $3<z<14$, \cite{Langeroodi24} found that dust must have appeared rapidly and that attenuation follows star formation on timescales shorter than $\sim$30\,Myr, favouring the scenario where dust is produced by supernovae (SN).
They found a linear correlation between A$_{\rm V}$ and the inverse sSFR (sSFR$^{-1}$, a proxy for the age of the galaxies) for galaxies with redshift from 3 to 14, and $\beta$ from -3 to 0.
Through a model, they show that the slope of this relation, $\gamma$, depends on several dust properties such as the efficiency of SN in producing dust, the fraction of dust destroyed by SN reverse and forward shocks, the fraction of dust removed from galaxy by feedback-driven outflows, dust geometry, and dust grain properties.
They obtained $\gamma=0.67\pm0.2$\,cm$^2$g$^{-1}$, 30 times lower than the value expected if the entire SN dust yield were preserved in the ISM, and had Milky-Way-like grain properties.
Here, we use our photometry-based estimates of A$_{\rm V}$, M$_\star$, and SFR, to compare the results obtained with thus two different methods.

\begin{figure*}[!htbp] 
 	\includegraphics[width=\columnwidth]{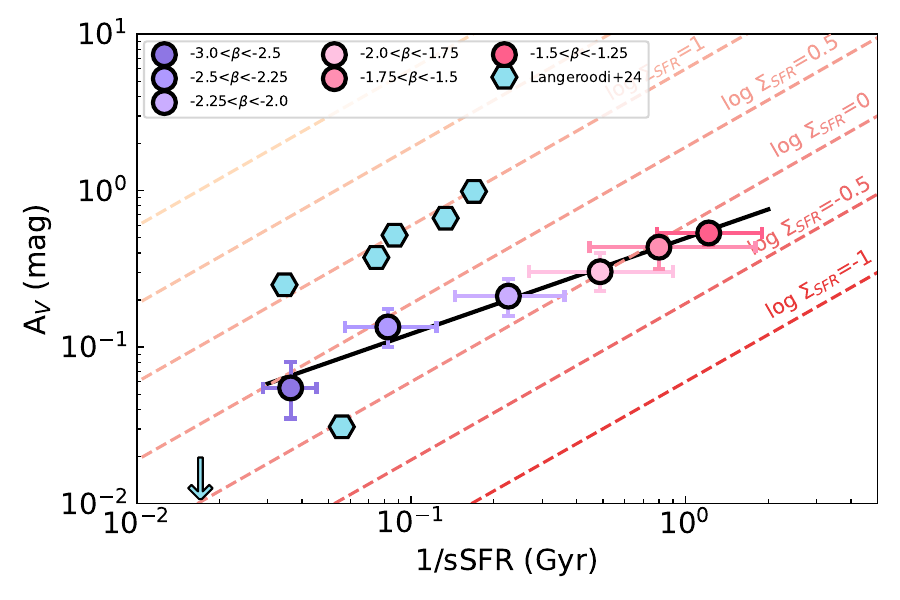}
 	\includegraphics[width=\columnwidth]{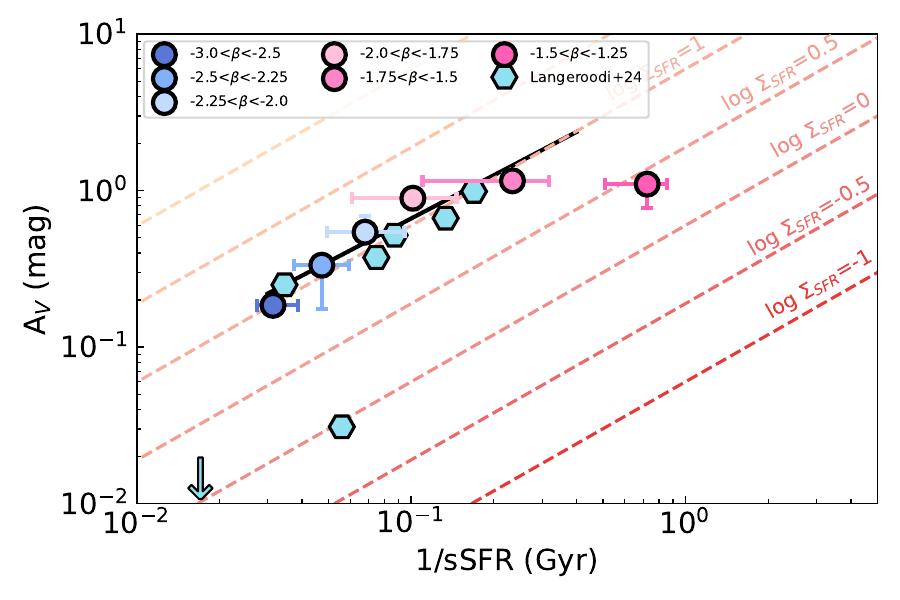}
  	\caption{\label{fig:avssfr} \textbf{Left:} A$_{\rm V}$ attenuation as a function of sSFR$^{-1}$. Galaxies with $6<z<12$ are binned by UV $\beta$ slope (circles). The black solid line is the linear fit of these binned points. Predictions of linear A$_{\rm V}$-sSFR$^{-1}$ relation at fixed SFR surface density from \cite{Langeroodi24} are indicated with coloured dashed lines. The stacked points obtained by \cite{Langeroodi24} are shown as cyan hexagons. \textbf{Right:} Same as in the left panel but using the A$_{\rm V}$ and sSFR$^{-1}$ obtained with the UV-NIR SED fit, that is without ALMA constraint. }
\end{figure*}

In Fig.~\ref{fig:avssfr}, we show the V-band attenuation as a function of the inverse sSFR for the whole sample ($6<z<12$) binned by UV slope $\beta$, similarly to \cite{Langeroodi24}.
Consistently with the results of the previous sections, we confirm a rapid dust production in early galaxies with non-negligible dust attenuation at sSFR$^{-1}>30$\,Myr.
Despite using different methods (spectroscopy vs photometry-based SED modelling), our data points also confirm the linear relation between A$_{\rm V}$ and sSFR$^{-1}$ found by \cite{Langeroodi24}, implying that dust production is driven by star formation activity and stellar mass build-up.

However, the intercept and slope of the relation obtained by our sample are lower and flatter, respectively:
\begin{equation}
\log A_{\mathrm{V}} = (0.61\pm0.05) \times \log \mathrm{sSFR}^{-1} - (0.30\pm0.03).
\end{equation}

First we checked if the flatter slope that we observe is due to the SED modelling by performing a mock analysis \citep[e.g.,][]{Buat15,Ciesla17,Boquien19}.
Basically, we ran \texttt{CIGALE} once on the JADES galaxies.
We took the best fit SED, convolved it in the JADES+ALMA set of filters, added noise to the photometry.
We ran again \texttt{CIGALE} on the mock photometry. 
In this way we can compare estimated parameters to the true ones used to build the mock photometry. 
From this test, we find that sSFR$^{-1}$ is recovered within errorbars with a systematic overestimate (60\%) and that A$_{\rm V}$ is systematically underestimated by -0.4\,mag. 
However, applying corrections derived from this mock analysis does not yield a slope consistent with the observations from \cite{Langeroodi24} (0.58$\pm$0.12).
Although we looked at the impact of dust temperature assumption in Sect.~\ref{tdust} and concluded that it does not affect strongly our conclusions for the $6<z<7$ redshift bin, studies predict that dust temperature keep increasing with redshift \citep[e.g.,][]{Sommovigo22,Mitsuhashi24}.
The temperature of 47\,K that we assumed in this work is consistent with the results and prediction of \cite{Sommovigo22} and \cite{Mitsuhashi24}.
However, it can reach up to 60-65\,K at $z>7$.
We thus test the impact of increasing the assumption on T$_{\rm dust}$ for galaxies in the $9<z<12$ to 60\,K (maximum value available in \cite{Schreiber18} library) and look at the impact of it on the slope of the A$_{\rm V}$-sSFR$^{-1}$ relation.
We obtain a slope of 0.63$\pm$0.05 and conclude that the assumption on T$_{\rm dust}$ is not causing the difference observed between our study and the results of \cite{Langeroodi24}.

We then test if the use of ALMA constraints is causing the discrepancy. 
Indeed, left panel Fig.~\ref{fig:comps} shows that A$_{\rm FUV}$ is lower when ALMA information is used in the SED modelling.
However, using A$_{\rm V}$, M$_{\star}$, and SFR obtained from the UV-to-NIR SED modelling run (Fig.~\ref{fig:comps}, right panel), our binned points are in perfect agreement with \cite{Langeroodi24}. 
We obtain a slope of 0.91$\pm$0.17 if we discard the reddest $\beta$ bin which seems to be discrepant and has a much lower statistics than the other bins.
Therefore, we conclude that the addition of ALMA constraints is at the origin of the different slopes between the spectroscopic sample of \cite{Langeroodi24} and our photometric one.
Our flatter slope increases even more the discrepancy with $\gamma$ obtained in Milky-Way conditions, but also points towards a relation between sSFR$^{-1}$ and the SFR surface density ($\Sigma_{\rm SFR}$
) for which we do not have access for our sample.

\section{\label{conclusions}Conclusions}

To place constraints on dust emission in galaxies at the epoch of reionization, we have used the A$^3$COSMOS (A$^3$GOODSS) database to stack all JADES galaxies at $6<z<12$ with available ALMA coverage in band 6 and 7. 
Focusing on the $6<z<7$ redshift bin, our results are summarized as follows:

\begin{itemize}

    \item ALMA stacks result in upper limits for galaxies with $M_{\mathrm{UV}}>$-19 and $\log M_\star<9$, and tentative detections at brighter UV magnitudes and higher stellar masses at $6<z<7$.

    \item Upper limits for UV faint galaxies are in agreement with predicted ALMA fluxes from UV-to-NIR SED fitting while ALMA fluxes are lower than predicted for UV bright $M_{\mathrm{UV}}<$-20 galaxies. For galaxies binned with stellar mass, the predicted fluxes without ALMA constraint and the ALMA fluxes are in agreement. However, for galaxies with $\log M_\star<9$, there is a tension between the ALMA derived upper limit (3$\sigma$) and the flux expected from scaling down the ALMA flux of the upper stellar mass bin. This is a hint at a possible lower dust to stellar mass ratio at these low stellar masses.

    \item The addition of the ALMA constraints for the SED modelling results in lower SFR (0.4\,dex), lower $A_{\rm FUV}$ (0.5\,dex) for galaxies with $\log M_\star>8$, as well as slightly higher stellar masses and lower $\beta$ for the same sources, implying older stellar population (25\% older on average).

    \item Using infrared constraints for the SED modelling, our $\beta$ vs $M_{\mathrm{UV}}$ is consistent with measurements from the literature, and in perfect agreement with the relation from \cite{Topping24}.

    \item Our measurements extend the $L_{\mathrm{IR}} - M_{\mathrm{UV}}$ relation down to $M_{\mathrm{UV}}\sim -19$ and hint at a possible breakdown of the relation at fainter UV magnitudes.

    \item When the ALMA constraints are used in the fitting, galaxies are consistent with the literature (REBELS and ALPINE) when placed on the IRX vs $\beta$ indicating similar dust composition and content than in these UV-selected samples. Our stack measurements computed in bins of stellar masses do not provide strong constraints on the IRX-$M_\star$ relation. However, the position of our UV bright bin ($M_{\mathrm{UV}}<-20$) is in good agreement with the relation from \cite{Bowler24}. Nevertheless, we find that a given stellar mass, the IRX decreases with increasing redshift.
\end{itemize}

We tested the impact of dust temperature assumption on our results and find that temperature assumptions between 36\,K and 47\,K yield similar $L_{\mathrm{IR}}$. 
The conclusions obtained with a dust temperature of 47\,K are still valid when changing the assumed dust temperature: In the IRX-$\beta$ diagram, our points are compatible with a \cite{Calzetti00} relation using $\beta_0=-2.3$ or the \cite{Reddy18} relation for starburst and SMC ($\beta_0=-2.6$). 
Regarding IRX-$M_\star$, even with a more accurate estimate of the dust temperature, it will be difficult to discriminate, within the errors, between the different relations from the literature in this $\beta$ range.

Combining the whole sample ($6<z<12$), we obtain the following results:
\begin{itemize}

    \item We did not find a clear evolution of $\beta$ with redshift for the galaxies of our sample and showed that this absence of variation is not due to the lower limit in $\beta_0$ imposed by our SED modelling procedure as only 5\% of galaxies at $z>9$ have the lowest value, and 15\% of the whole sample at $z>6$. Considering that the median $M_{\mathrm{UV}}$ of our sample is -17.24$_{-0.62}^{+0.54}$, this non evolution is consistent with the results of \cite{Topping24} at similar UV magnitude range.

    \item We provided upper limits on the IR luminosity of NIRCam-selected galaxies at $z>7$, and show that detected sample from the literature (REBELS, SERENADE, and individual detections) are not representative of the bulk of galaxies at these redshifts. 

    \item We confirmed a linear relation between the V-band attenuation and the inverse sSFR showing a non-negligible attenuation at sSFR$^{-1}>30$\,Myr and a flatter slope than obtained from \cite{Langeroodi24} from a spectroscopic sample, due to the use of ALMA constraints in our physical properties measurements.

\end{itemize}

Our results focused on $6<z<7$ galaxies indicate that the dust properties at the Epoch of Reionization are mostly similar to lower redshift galaxies, implying that a significant dust production must have already happened by $z\sim6.5$.
However, the breakdown observed in the $L_{\rm IR}$ versus $M_{\rm UV}$ relation at $M_{\rm UV}\geq$-19 is possible hint at a different regime in the UV faint galaxies. 
These galaxies have on average low masses ($\log M_\star\sim8$) resulting in shallower gravitational potential that could allow metals to escape, thus resulting in less dust.
Or it could be also be due to the lower SFR measured in these sources ($\sim$0.5-2\,M$_{\odot}$yr$^{-1}$) resulting in a lower dust production.
Nevertheless, the rapid build-up concluded from the $6<z<7$ galaxies is also confirmed by the non-negligible amount of attenuation observed at sSFR$^{-1}>30$\,Myr for galaxies with $6<z<12$.

\begin{acknowledgements}
LC warmly thanks A.~Ferrara, J.~Lewis, M.~Trebitsch, and D.~Elbaz for insightful discussions that improved greatly this work.
The authors thank the JADES team for the huge work and effort put in the preparation, observation, and production of the data used in this work.
This work received support from the French government under the France 2030 investment plan, as part of the Initiative d’Excellence d’Aix-Marseille Université – A*MIDEX AMX-22-RE-AB-101.
This project has received financial support from CNRS and CNES through the MITI interdisciplinary programs.
SA gratefully acknowledges the Collaborative Research Center 1601 (SFB 1601 sub-project C2) funded by the Deutsche Forschungsgemeinschaft (DFG, German Research Foundation) – 500700252. 
LC and MB acknowledges the funding of the French Agence Nationale de la Recherche for the project iMAGE (grant ANR-22-CE31-0007).
NIRCam was built by a team at the University of Arizona (UofA) and Lockheed Martin's Advanced Technology Center, led by Prof. Marcia Rieke at UoA.
ES acknowledges funding from the European Research
Council (ERC) under the European Union’s Horizon 2020 research and innovation
programme (grant agreement No. 694343).
This work is based on observations made with the NASA/ESA/CSA James Webb Space Telescope. 
The data were obtained from the Mikulski Archive for Space Telescopes at the Space Telescope Science Institute, which is operated by the Association of Universities for Research in Astronomy, Inc., under NASA contract NAS 5-03127 for JWST. 
MB gratefully acknowledges support from the ANID BASAL project FB210003 and from the FONDECYT regular grant 1211000. This work was supported by the French government through the France 2030 investment plan managed by the National Research Agency (ANR), as part of the Initiative of Excellence of Université Côte d'Azur under reference number ANR-15-IDEX-01.
\end{acknowledgements}
\bibliographystyle{aa}
\bibliography{jades_stack_alma}

\end{document}